\documentclass[epj]{svjour}
\usepackage{graphicx,subfigure}
\usepackage{textcomp,color}
\usepackage{bm}        
\usepackage{amssymb}   
\usepackage{mathrsfs}
\usepackage{amsmath}

\begin{document}
\title{Extreme values of elastic strain and energy in sine-Gordon multi-kink collisions}
\subtitle{}
\author{Aliakbar Moradi Marjaneh\inst{1,}\thanks{\emph{Email:} moradimarjaneh@gmail.com} \and Alidad Askari\inst{2} \and Danial Saadatmand\inst{3} \and Sergey V. Dmitriev\inst{4,5} 
}                     
\offprints{}          
\institute{Department of Physics, Quchan Branch, Islamic Azad University, Quchan, Iran \and Department of Physics, Ferdowsi University of Mashhad, 91775-1436 Mashhad, Iran \and Department of Physics, University of Sistan and Baluchestan, Zahedan, Iran \and Institute for Metals Superplasticity Problems RAS, Khalturin Street 39, 450001 Ufa , Russia \and National Research Tomsk State University, Lenin Ave. 36, 634050 Tomsk, Russia}
\date{Received: date / Revised version: date}
%
\abstract{
In our recent study the maximal values of kinetic and potential energy densities that can be achieved in the collisions of $N$ slow kinks in the sine-Gordon model were calculated analytically (for $N=1,2$, and 3) and numerically (for $4\le N\le 7$). However, for many physical applications it is important to know not only the total potential energy density but also its two components (the on-site potential energy density and the elastic strain energy density) as well as the extreme values of the elastic strain, tensile (positive) and compressive (negative). In the present study we give (i) the two components of the potential energy density and (ii) the extreme values of elastic strain. Our results suggest that in multi-soliton collisions the main contribution to the potential energy density comes from the elastic strain, but not from the on-site potential. It is also found that tensile strain is usually larger than compressive strain in the core of multi-soliton collision. 
} 
\maketitle
\section{Introduction}
\label{intro}

Solitons (or solitary waves) are present in a wide range of physics such as optics \cite{Optics1,Optics2}, superconducting Josephson junction arrays \cite{JJ1,JJ2,JJ3}, particle and nuclear physics \cite{Manton,Weigel}, condensed matter physics \cite{Bishop,Phys,Ferro1,Ferro3,Ferro2,Buijnsters,Wrinklon,Yamal}, and many others \cite{Belova,BraunKivshar,BookSGE,nSGE}. They are robust with respect to small perturbations, they can travel long distances maintaining their shape, and survive collisions with each other. That is why solitons are considered to be very efficient in transferring energy, momentum, mass, electric charge and other physical quantities, depending on the application.

Multi-soliton solutions to the sine-Gordon (sG) equation have been derived analytically \cite{Perring,Hirota} and analyzed by many authors \cite{Caudrey,GK1999,Wazwaz,Ekomas1,Ekomas2,GLL2015,DKK2008,DKS2001,Mirosh,Javidan}. Colliding solitons can produce high energy density spots and for many applications it is important to know how large the energy density can be in multi-soliton collisions. Recently we have addressed this issue in the sG, $\phi^4$, and $\phi^6$ models \cite{DanialPRD,Aliakbar,NewJHEP}. Basic soliton solutions to sG equation are the kink, which interpolates between nearest wells of the periodic on-site potential, and the breather (or bion), which is a bound state of a kink and antikink \cite{Perring}. It was found that maximal energy density that can be observed in collision of $N$ slowly moving kinks/antikinks is proportional to $N^2$ in all three models, while total energy of the system is proportional to $N$. Such a high energy density can be achieved only if the kinks and antikinks approach the collision point in an alternating array, where each soliton has nearest neighbors of the opposite topological charge and thus, all solitons attract their nearest neighbors. Interestingly, in the case of sG equation, when $N$ is odd (even) the maximal energy density is in the form of potential (kinetic) energy with kinetic (potential) energy density being zero. In the $\phi^4$ and $\phi^6$ models, for odd (even) $N$ the maximal energy density is mainly in the form of potential (kinetic) energy.

Solitons in non-integrable models such as $\phi^4$ model can support kink's internal vibrational modes \cite{Rajaraman,KPCP}, which make soliton dynamics richer than in the integrable sG model \cite{BraunKivshar,BookSGE,Quintero,phi6a,phi6b,GLL2015,phi8,sinhdeformed,domainwall,Khare,KL2017,JavidanPRE}. This is so because the kink's translational and vibrational modes can exchange energy. The effect of the non-integrability in the maximal energy density has been investigated in \cite{Aliakbar}. It was shown that the kink's internal mode can increase or decrease the maximal energy density in multi-kink collisions.

In the present study we revisit the results presented in \cite{DanialPRD} aiming to provide more information useful for physical applications. First of all, note that in the Klein-Gordon equations, including sG equation, potential energy consists of two parts, the energy related to the on-site potential and the elastic strain energy. In the published work \cite{DanialPRD} these energies were not separated, but they have very different physical nature and it is important to know how do they contribute to the total potential energy. Secondly, from the physical standpoint it is important to know not only the value of the maximal elastic strain energy density in multi-kink collisions, but also the value of the maximal tensile and compressive strain because the response of media to strain can differ qualitatively depending on the sign of strain. For instance, in the application of the sG equation to the dislocation theory \cite{FK}, the interatomic bonds can break if tensile strain is above a threshold value. Thus, the main goal of the present study is to extend the work \cite{DanialPRD} by the discussion of the extreme values of the two parts of the potential energy density and the extreme values of the tensile and compressive elastic strain in multi-kink collisions.

The paper is organized as follows. In Sec. \ref{Sec:II} the problem to solve is described. In Sec. \ref{Sec:III}, analytically (for $N\le 3$) and by integrating numerically the sG equation of motion (for $N>3$), we estimate the extreme values of energy density components and strain observed in the collision of $N$ slowly moving kinks and antikinks. The key results of the present study are summarized in Sec.~\ref{Sec:V}.

\section {General remarks} \label{Sec:II}
The equation of motion of the sG model in (1+1) dimension is
\begin{equation}\label{SG}
\phi_{tt} - \phi_{xx}+\sin\phi = 0,
\end{equation}
where $\phi_{tt}$  and $\phi_{xx}$ are the second order derivative of $\phi(x,t)$ scalar field with respect to the time and space, respectively. This equation has an exact solution which is known as kink (antikink)
\begin{equation}\label{Kink}
\phi_{k}(x,t) = \pm4\arctan{\exp[\delta(x-x_0-Vt)]}, 
\end{equation}
where $x_0$ is the initial position of the kink, $0\le V<1$ is the kink velocity and $\delta =1/\sqrt{1 -V^2}$. Positive (negative) sign in Eq.~(\ref{Kink}) is related to the kink (antikink). The total energy of the scalar field is defined by
\begin{equation}\label{Energy}
U=\int\limits_{-\infty}^{\infty} \Big[ \frac{1}{2}\phi_t^2 +\frac{1}{2}\phi_x^2 +(1-\cos\phi) \Big]dx \, .
\end{equation}
Three integrands in Eq.~(\ref{Energy}) describe the three contributions to the total energy density of the sG field,
\begin{eqnarray}\label{Edensity}
u(x,t)&=& k(x,t)+e(x,t)+p(x,t) \, ,
\end{eqnarray}
where
\begin{equation}\label{Edensityk}
k(x,t)=\frac{1}{2}\phi_t^2,
\end{equation}
\begin{equation}\label{Edensitye}
e(x,t)=\frac{1}{2}\phi_x^2,
\end{equation}
\begin{equation}\label{Edensityp}
p(x,t)=1-\cos\phi,
\end{equation}
are the kinetic energy density, the elastic strain density and the on-site potential energy density, respectively.

Elastic strain of the system which is positive (negative) for tension (compression) can be defined as follows
\begin{eqnarray}\label{Strain}
\varepsilon(x,t)=\phi_x.
\end{eqnarray}

By substituting an exact solution of the sG field from Eq.~(\ref{Kink}) into Eq.~(\ref{Energy}) the total energy of the kink can be obtained 
\begin{equation}\label{KBEnergy} 
U=8\delta.
\end{equation}
In this study only slow kinks ($|V| \ll 1$) are analyzed, so that $\delta \approx 1$. Then, the total energy for one slow kink is approximately $U \approx 8$, and for $N$ slow kinks/antikinks it is about $8N$.

To carry out numerical stimulations, the mesh $x_n = nh$, $t_j = j\tau$ is introduced, where $h$ is the lattice spacing, $\tau$ is the time step, $n=0,\pm1,\pm2,...$ and $j =-3,-2,-1,0,1,...$. The discrete form of Eq.~(\ref{SG}) is considered in the form
\begin{eqnarray}\label{SG discrete}
&&\Big(\frac{d^2\phi_n}{dt^2}\Big)_j - \frac{1}{h^2}(\phi_{n-1,j} -2\phi_{n,j}+\phi_{n+1,j}) \nonumber \\
&&+\frac{1}{12h^2}(\phi_{n-2,j}-4\phi_{n-1,j} +6\phi_{n,j}-4\phi_{n+1,j}+\phi_{n+2,j})\nonumber \\
&&+\sin(\phi_{n,j})  = 0,
\end{eqnarray}
where the second derivative with respect to time at $j$-th time step is discretized as follows
\begin{eqnarray}\label{SecDeriv}
&&\Big(\frac{d^2\phi_n}{dt^2}\Big)_j  \nonumber \\
&&= \frac{11\phi_{n,j+1}-20\phi_{n,j} +6\phi_{n,j-1}+4\phi_{n,j-2}-\phi_{n,j-3}}{12\tau^2}.
\end{eqnarray}
This discretization method is fourth order accurate in space \cite{BraunKivshar,DanialPRD}. Note that in order to increase the accuracy of the integration over the temporal variable the four-step scheme is used with the initial conditions specified for steps $j = -3, -2, -1$, and 0. Substituting Eq.~(\ref{SecDeriv}) into Eq.~(\ref{SG discrete}) one can express $\phi_{n,j+1}$ in terms of grid function values at $j - 3$, $j - 2$, $j - 1$, and $j$ time steps, thus obtaining an explicit scheme with the accuracy of $O(\tau^4)$. The simulations were conducted for the temporal step $\tau=0.005$ and it was checked that further decrease of $\tau$ does not affect the numerical results noticeably. The results for the spatial steps $h=0.1$ and $h=0.05$ are compared to see the convergence with respect to this parameter of the numerical scheme. No instability problems were observed for chosen parameters of the numerical scheme.

Fixed boundary conditions are used. The size of the computational cell of $20,000$ points was sufficiently large to exclude the effect of small amplitude waves reflected from the boundaries on the kink collision dynamics. 

In our numerical simulations for $N>3$ the initial conditions are specified with the help of the exact kink/antikink solution Eq.~(\ref{Kink}) in a way that the solitons initially do not overlap (initial distance between them is not less than 12). All solitons in the initial configuration have opposite topological charges with the neighbors. This ensures attractive forces between all neighboring  solitons, which is a necessary condition for all of them to collide at one point and produce the highest possible energy density (see Refs.~\cite{DanialPRD,Aliakbar}). To achieve collision of all of $N$ solitons at one point, one has to set particular values of initial velocities of solitons and initial distances between them. The choice of these parameters will be discussed for each number of colliding solitons in Sec.~\ref{Sec:III}.
In our simulations we set $t=0$ at the collision point and show the results in the time domain near this point. The simulation run is typically within the range $200<t<200$.

\section {Extreme values of elastic strain and energy in multi-kink collisions} \label{Sec:III}

In this Section, we firstly calculate analytically the exact values of the maximal energy densities and maximal elastic strain for single standing kink. Afterwards, the same quantities are obtained analytically from the exact separatrix solutions to sG for $N=2$ and $N=3$ kinks and antikinks colliding with the smallest possible velocity.  Then the same quantities are estimated numerically for $N=4$, 5, 6 and 7 slowly moving kinks and antikinks colliding at one point. 

In this study we analyze collisions between slowly moving kinks and antikinks. It is clear that fast collisions will produce higher energy density in comparison to slow collisions because faster kinks have smaller width and a greater contribution from the kinetic energy density. In the relativistic limit $|V|\rightarrow 1$, kink width vanishes and its energy diverges, as follows from Eq.~(\ref{KBEnergy}), making comparison of maximal energy densities for different collision scenarios meaningless. On the other hand, for slow kinks with $|V|\ll 1$, kink energy is a weak function of $V$ and one can compare maximal energy densities for different number of colliding solitons without ambiguity.
\subsection{The case of single kink} \label{Sec:N1}
Substitution Eq.~(\ref{Kink}) with $V=0$ in Eq.~(\ref{Edensity}) yields the next maximal values of the energy densities of standing kink (antikink):
\begin{equation}\label{umax1}
u^{(1)}_{\max}=4,
\end{equation}
\begin{equation}\label{kmax1}
k^{(1)}_{\max}=0,
\end{equation}
\begin{equation}\label{emax1}
e^{(1)}_{\max}=2,
\end{equation}
\begin{equation}\label{pmax1}
p^{(1)}_{\max}=2.
\end{equation}
In this case the total maximal energy density, $u^{(1)}_{\max}$, is equal to the sum of $k^{(1)}_{\max}$, $e^{(1)}_{\max}$, and $p^{(1)}_{\max}$; but in the following we will see that this is not so for $N>1$.

By substituting Eq.~(\ref{Kink}) with $V=0$ into Eq.~(\ref{Strain}) the extreme values of strain are extracted as
\begin{equation}\label{epsmax1}
   \varepsilon^{(1)}_{\max}= 2, \quad \varepsilon^{(1)}_{\min}=-2,
\end{equation}
where the positive (negative) value answers the kink (antikink).

\subsection{Two-kink collisions} \label{Sec:N2}

\begin{figure*}
\center
\resizebox{0.8\textwidth}{!}{%
  \includegraphics{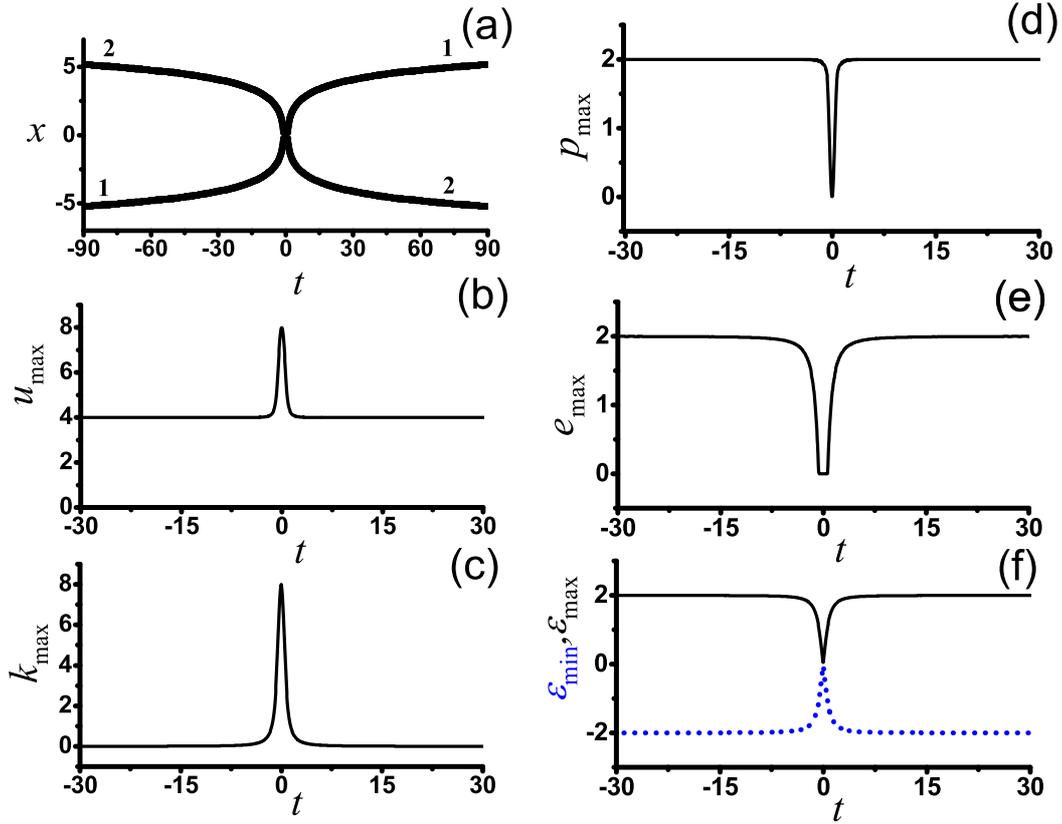}
}
\caption{The plot corresponds to the kink-antikink separatrix solution Eq.~(\ref{Kak}). (a)~ Trajectories of the soliton cores in the $(t,x)$-plane shown by plotting the regions where total energy density $u(x,t)>2$. (b-e)~Maximal over spatial coordinate values of energy densities $u$, $k$, $p$ and $e$ as the functions of time, respectively. (f) Maximal (black solid line) and minimal (blue dotted line) over spatial coordinate values of elastic strain $\varepsilon$.}
\label{fig1}       
\end{figure*}

The separatrix two-soliton solution of the sG model is given by \cite{DKK2008}
\begin{equation}\label{Kak}
 \phi_{k\bar k}(x,t) =-4\arctan\frac{t}{\cosh x}.
\end{equation}
This solution describes the kink and antikink that after the collision at $t=0$ move apart and their velocities vanish as $t\rightarrow \infty$. Thus, this solution represents the kink-antikink collision with the smallest possible velocity. The solution and its parameters are shown in Fig.~\ref{fig1} for the times domain close the collision point. In (a) the regions of the ($t,x$)-plane with the total energy density $u>2$ are shown to reveal the cores of the solitons. Integrable sG equation describes purely elastic kink-antikink collision. From (b) to (f), the plots show the maximal over $x$ values of $u$, $k$, $p$, $e$, and $\varepsilon$ as the functions of time, respectively. To calculate the exact values of the maximal energy densities we substitute Eq.~(\ref{Kak}) into Eq.~(\ref{Edensity}), then we have
\begin{equation}\label{u2}
 u(x,t)=k(x,t)+e(x,t)+p(x,t),
\end{equation}
where
\begin{equation}\label{k2}
    k(x,t)=\frac{8{\rm sech}^2x}{(1+t^2{\rm sech}^2x)^2},
\end{equation}
\begin{equation}\label{e2}
    e(x,t)=\frac{8t^2{\rm sech}^2x\tanh ^2x}{(1+t^2{\rm sech}^2x)^2},
\end{equation}
\begin{equation}\label{p2}
    p(x,t)=1-\cos[4\arctan(t{\rm sech}x)].
\end{equation}
The maximal values of the energy densities for the two-soliton collision are
\begin{equation}\label{umax2theory}
   u^{(2)}_{\max}=8,
\end{equation}
\begin{equation}\label{kmax2theory}
   k^{(2)}_{\max}=8,
\end{equation}
\begin{equation}\label{emax2theory}
   e^{(2)}_{\max}=2,
\end{equation}
\begin{equation}\label{pmax2theory}
   p^{(2)}_{\max}=2.
\end{equation}
By substituting Eq.~(\ref{Kak}) into Eq.~(\ref{Strain}) the elastic strain and its extreme values are extracted as
\begin{equation}\label{eps2}
\varepsilon(x,t)=\frac{-4t{\rm sech}x\tanh x}{1+t^2{\rm sech}^2x},
\end{equation}
\begin{equation}\label{epsmax2theory}
  \varepsilon^{(2)}_{\max}=2 , \quad \varepsilon^{(2)}_{\min}=-2.
\end{equation}
%
\subsection{Three-kink collisions } \label{Sec:N3}
\begin{figure*}
\center
\resizebox{0.75\textwidth}{!}{%
  \includegraphics{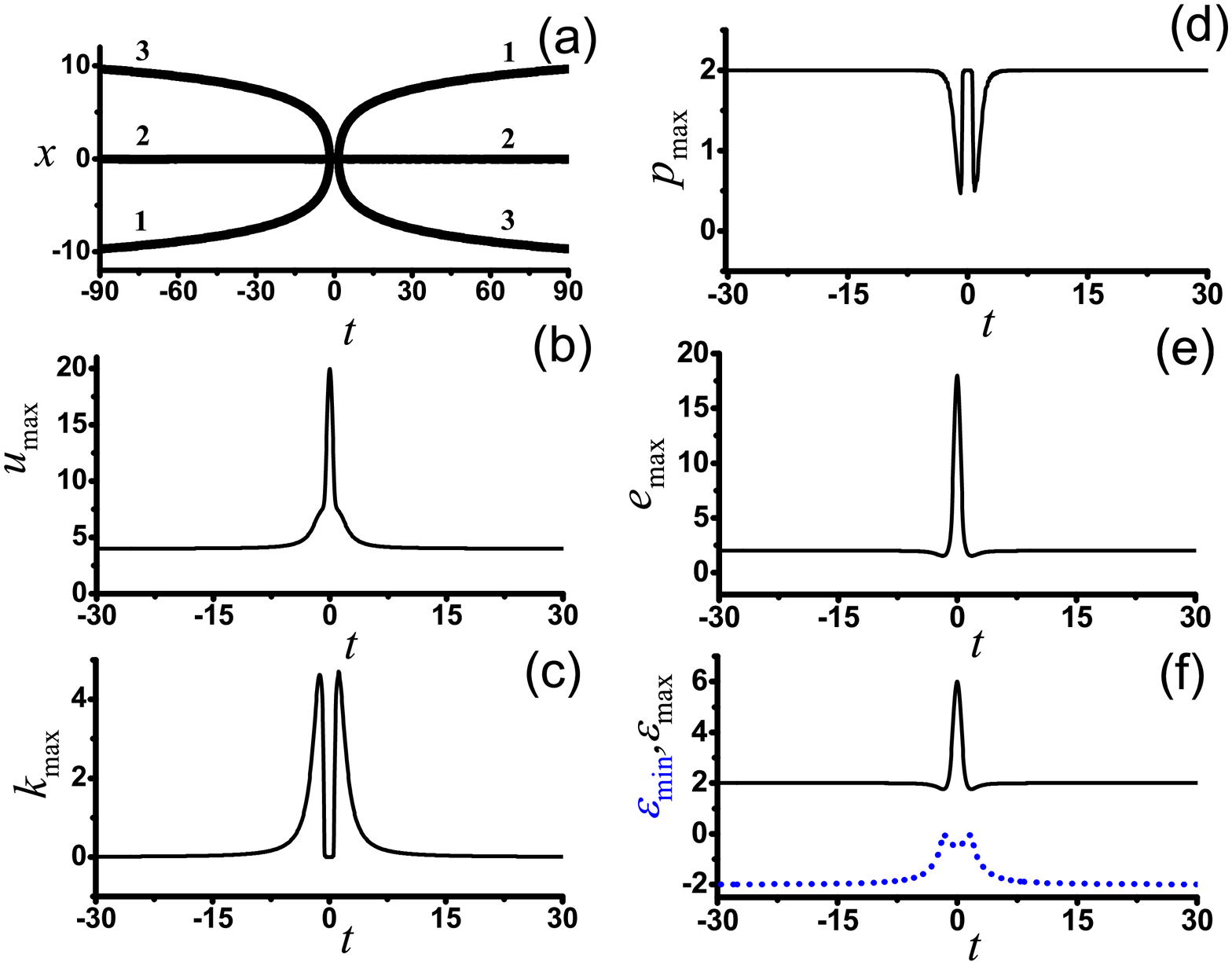}
}
\caption{Same as in Fig.~\ref{fig1}, but for the exact kink-antikink-kink separatrix solution Eq.~(\ref{Kakk}).}
\label{fig2}
\end{figure*}
The separatrix three-soliton solution of the sG model is \cite{DKK2008}
\begin{align}\label{Kakk}
\phi(x,t)=4\arctan[\exp(x)]
+4\arctan\frac{x \cosh x-t^2\sinh x}{t^2+\cosh ^2x}.
\end{align}
This solution describes the antikink standing at $x=0$ and two kinks that after the collision with the antikink at $t=0$ move apart and their velocities vanish as $t\rightarrow0$. This solution describes three-soliton collision with the smallest possible velocity. The solution is plotted in Fig.~\ref{fig2} similarly to Fig.~\ref{fig1}. Since unperturbed sG equation is considered, there is no energy exchange between colliding solitons.

By substituting Eq.~(\ref{Kakk}) into Eq.~(\ref{Edensity}) we find three components of the total energy density as follows:
\begin{equation}\label{k3}
    k(x,t)=\frac{8 [\frac{2 t \sinh x}{t^2+\cosh ^2x} +\frac{2 t (x \cosh x-t^2 \sinh x)}{(t^2+\cosh ^2 x)^2}]^2}{[\frac{(x \cosh x-t^2 \sinh x)^2}{(t^2+\cosh ^2 x)^2}+1]^2},
\end{equation}
\begin{align}\label{e3}
    e(x,t)&=\frac{1}{2}[\frac{4\exp(x)}{\exp(2x)+1}\nonumber\\
 &-\frac{8 \sinh x\cosh x (x \cosh x-t^2 \sinh x)}{(t^2+\cosh ^2 x)^2 (\frac{(x \cosh x-t^2 \sinh x)^2}{(t^2+\cosh ^2 x)^2}+1)}\nonumber\\
 & + \frac {4 (-t^2 \cosh x +x\sinh x + \cosh x)} {(t^2 + \cosh^2 x) (\frac {(x\cosh x - 
             t^2\sinh x)^2} {(t^2 + \cosh^2 x)^2} + 1)}]^2.
\end{align}
\begin{align}\label{p3}
    p(x,t)&=1-\cos[4\arctan(\exp(x))\nonumber\\
 &+4\arctan\frac{x\cosh x-t^2 \sinh x}{t^2+\cosh ^2 x}],
\end{align}
The exact values of the maximal energy densities are
\begin{equation}\label{umax3theory}
   u^{(3)}_{\max}=20,
\end{equation}
\begin{equation}\label{kmax3theory}
   k^{(3)}_{\max}=4.80915,
\end{equation}
\begin{equation}\label{emax3theory}
   e^{(3)}_{\max}=18,
\end{equation}
\begin{equation}\label{pmax3theory}
   p^{(3)}_{\max}=2.
\end{equation}
By substituting Eq.~(\ref{Kakk}) into Eq.~(\ref{Strain}) the elastic strain is extracted as
\begin{align}\label{eps3}
\varepsilon(x,t)&=\frac{4\exp(x)}{\exp(2x)+1}\nonumber\\
 &-\frac{8 \sinh x\cosh x (x \cosh x-t^2 \sinh x)}{(t^2+\cosh ^2 x)^2 (\frac{(x \cosh x-t^2 \sinh x)^2}{(t^2+\cosh ^2 x)^2}+1)}\nonumber\\
 & + \frac {4 (-t^2 \cosh x +x\sinh x + \cosh x)} {(t^2 + \cosh^2 x) (\frac {(x\cosh x - 
             t^2\sinh x)^2} {(t^2 + \cosh^2 x)^2} + 1)},
\end{align}
with the maximum and minimum values
\begin{equation}\label{epsmax3theory}
   \varepsilon^{(3)}_{\max}=6 , \quad \varepsilon^{(3)}_{\min}=-2.
\end{equation}
\subsection{Four-kink collisions} \label{Sec:N4}
\begin{figure*}
\center
\resizebox{0.75\textwidth}{!}{%
  \includegraphics{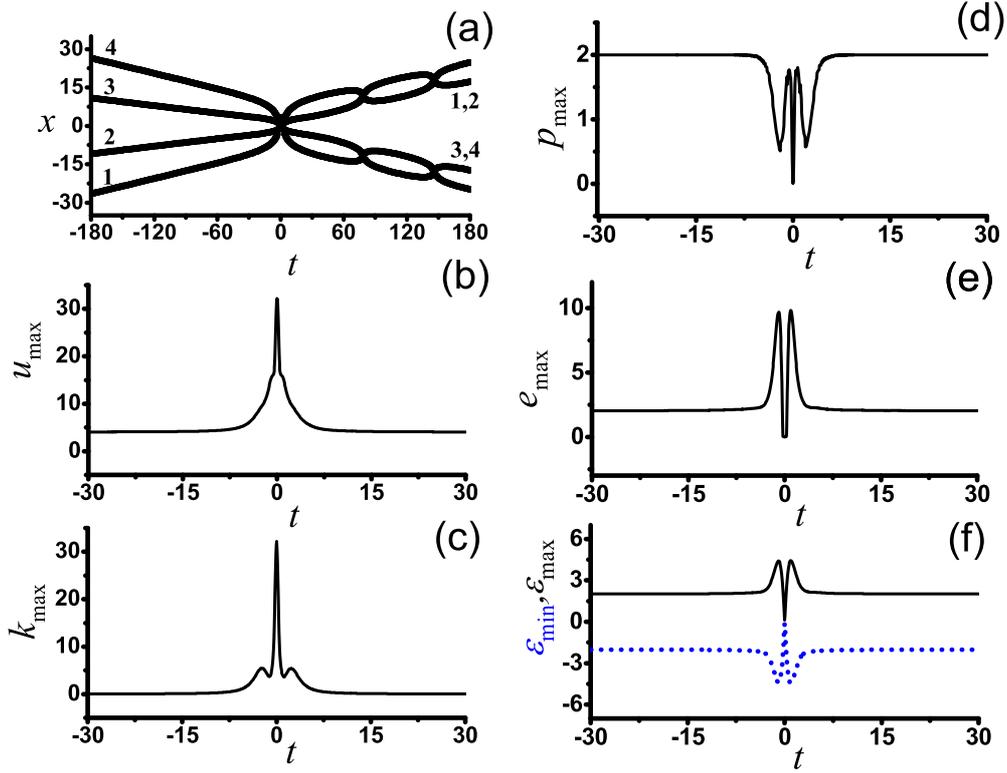}
}
\caption{Numerical results for the four-kink collision. Solitons 1 and 3 are kinks, while 2 and 4 are antikinks. Initial conditions are set by combination of well separated solitons specified by Eq.~(\ref{Kink}) with the initial coordinates $x_1=-x_4=-22.63887$, $x_2=-x_3=-9$ and initial velocities $V_1=-V_4=0.1$, $V_2=-V_3=0.05$.} 
\label{fig3}
\end{figure*}
In the initial configuration shown in Fig.~\ref{fig3}~(a), solitons 1 and 3 are the kinks, while 2 and 4 are the antikinks. They collide at one point provided that their initial positions and velocities
are chosen as follows: $x_1=-x_4=-22.63887$, $V_1=-V_4=0.1$, $x_2=-x_3=-9$ and $V_2=-V_3=0.05$. In this case (and in all other cases with $N>3$), sG equation is integrated numerically and the discreteness of the numerical scheme produces a sort of perturbation. This leads to merger of kinks-antikink pairs into bound states called breathers or bions.

We obtained from Fig.~\ref{fig3}~(b-f) for the collision of five solitons the following results: 
\begin{equation}\label{umax4}
u^{(4)}_{\max}\approx 32,
\end{equation}
\begin{equation}\label{kmax4}
k^{(4)}_{\max}\approx 32,
\end{equation}
\begin{equation}\label{emax4}
e^{(4)}_{\max}\approx 10,
\end{equation}
\begin{equation}\label{pmax4}
p^{(4)}_{\max}\approx 2,
\end{equation}
\begin{equation}\label{epsmax4}
\varepsilon^{(4)}_{\max}\approx 4.5, \quad \varepsilon^{(4)}_{\min}\approx -4.5.
\end{equation}
More precisely, for $h=0.1$ the largest energy density we could obtain by varying the parameter $x_{1}=-x_{4}$ (at fixed other parameters) was $u^{(4)}_{\max}=32.21$, while for $h=0.05$ it was $u^{(4)}_{\max}=32.05$.

A note should be made on how the kinks and antikinks are numbered after the collisions in Fig.~\ref{fig3}~(a) and in other similar plots. Considered system is nearly integrable, since it is weakly perturbed by the discreteness of the numerical scheme. In the integrable system solitons after collision fully recover their shapes and properties, so that, e.g., their velocities after the collision are the same as before the collision. For our nearly integrable system we assume that the order of kinks/antikinks after the collision is reversed (as it must be in an integrable system), and we number them accordingly. The kink-antikink pairs that merge into a breather have two numbers separated by a comma.
\subsection{Five-kink collisions} \label{Sec:N5}
\begin{figure*}
\center
\resizebox{0.75\textwidth}{!}{%
  \includegraphics{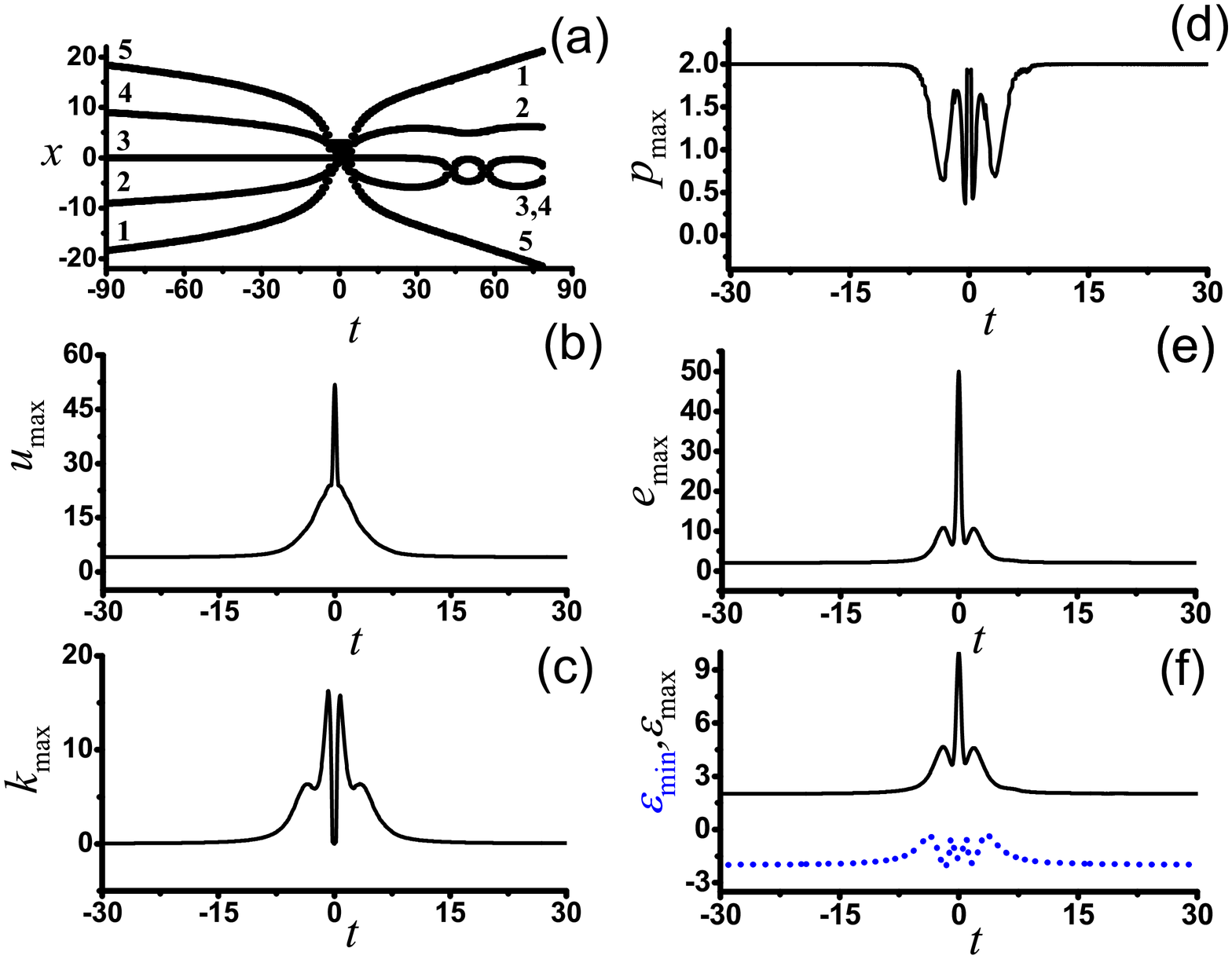}
}
\caption{Numerical results for the five-kink collision. Solitons 1, 3 and 5 are kinks, while 2 and 4 are antikinks. Initial conditions are set by combination of well separated solitons specified by Eq.~(\ref{Kink}) with the initial coordinates
$x_1=-x_5=-24.376549$, $x_2=-x_4=-12.0$, $x_3=0$, and initial velocities $V_1=-V_5=0.05$, $V_2=-V_4=0.025$, $V_3=0$.}
\label{fig4}
\end{figure*}

In Fig.~\ref{fig4}~(a) solitons 1, 3, and 5 are kinks and 2 and 4 are antikinks. The kink 3 is located at the
origin with initial condition, $x_3=0$ and $V_3=0$.  The velocities of the antikinks 2 and 4 are  $V_2=-V_4=0.025$ and they are initially  located at the positions $x_2=-x_4=-12.0$.  By symmetry the solitons 2, 3, and 4 collide at one point. For the kinks 1 and 5 we take
two times larger velocities $V_1=-V_5=0.05$ and choose their initial coordinates
to achieve the collision of five solitons at one point. This happens for
$x_1=-x_5=-24.376549$. As it can be seen from Fig.~\ref{fig4}(b-f), when five solitons collide,
\begin{equation}\label{umax5}
u^{(5)}_{\max}\approx 52,
\end{equation}
\begin{equation}\label{kmax5}
k^{(5)}_{\max}\approx 17,
\end{equation}
\begin{equation}\label{emax5}
e^{(5)}_{\max}\approx 52,
\end{equation}
\begin{equation}\label{pmax5}
p^{(5)}_{\max}\approx 2,
\end{equation}
\begin{equation}\label{epsmax5}
\varepsilon^{(5)}_{\max}\approx 10, \quad \varepsilon^{(5)}_{\min}\approx -2.
\end{equation}
More precisely, the maximal energy density is $u^{(5)}_{\max}=50.93$ for $h=0.1$ and $u^{(5)}_{\max}=51.85$ for $h=0.05$.
\subsection{Six-kink collisions} \label{Sec:N6}
\begin{figure*}
\center
\resizebox{0.75\textwidth}{!}{%
  \includegraphics{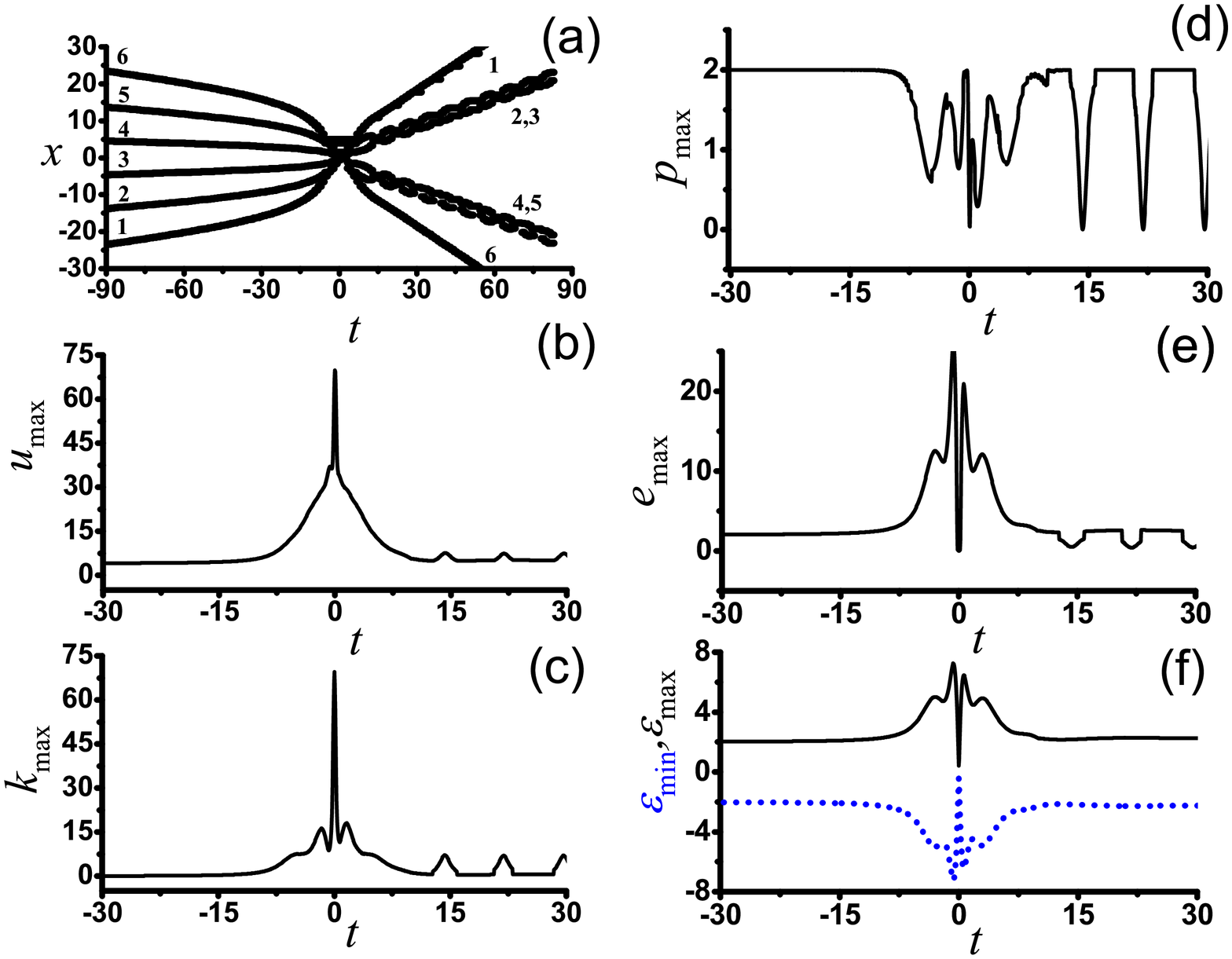}
}
\caption{Numerical results for the six-kink collision. Solitons 1, 3 and 5 are kinks, while 2, 4 and 6 are antikinks. Initial soliton coordinates and velocities are  $x_1=-x_6=-34.90395$, $V_1=-V_6=0.1$, $x_2=-x_5=-19.37864$, $V_2=-V_5=0.05$, $x_3=-x_4=-7$, and $V_3=-V_4=0.025$.
}
\label{fig5}
\end{figure*}

In Fig.~\ref{fig5} the solitons 1, 3 and 5 are kinks and 2, 4 and 6 are antikinks. The initial positions and velocities to provide the collision of all solitons at one point are $x_1=-x_6=-34.90395$, $V_1=-V_6=0.1$, $x_2=-x_5=-19.37864$, $V_2=-V_5=0.05$, $x_3=-x_4=-7$, $V_3=-V_4=0.025$. The maximal of energy densities and maximal compressive and tensile strains are found to be
\begin{equation}\label{umax6}
u^{(6)}_{\max}\approx 72,
\end{equation}
\begin{equation}\label{kmax6}
k^{(6)}_{\max}\approx 72,
\end{equation}
\begin{equation}\label{emax6}
e^{(6)}_{\max}\approx 25,
\end{equation}
\begin{equation}\label{pmax6}
p^{(6)}_{\max}\approx 2,
\end{equation}
\begin{equation}\label{epsmax6}
\varepsilon^{(6)}_{\max}\approx 7, \quad \varepsilon^{(6)}_{\min}\approx -7.
\end{equation}

\subsection{Seven-kink collisions} \label{Sec:N7}
\begin{figure*}
\center
\resizebox{0.74\textwidth}{!}{%
  \includegraphics{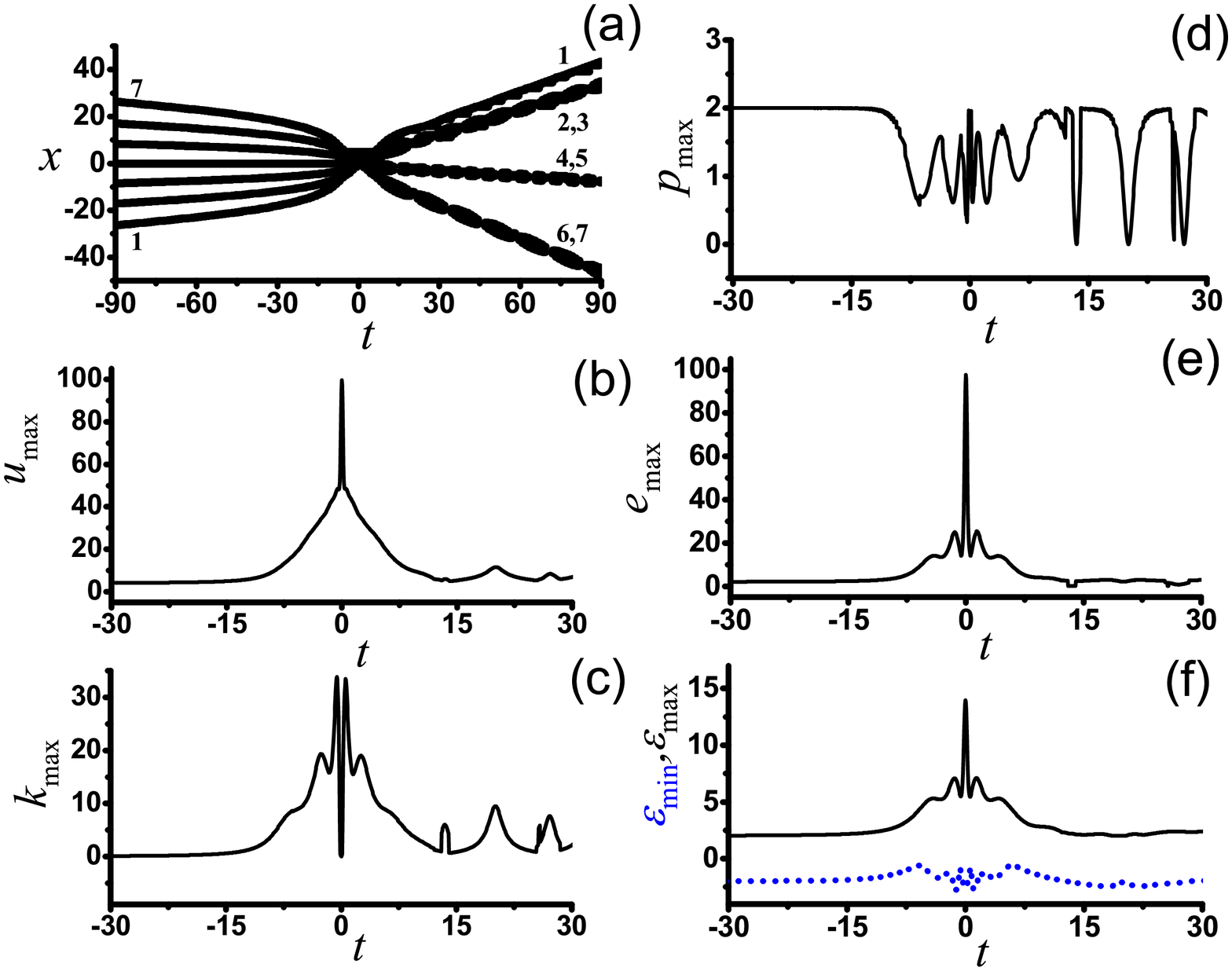}
}
\caption{Numerical results for the seven-kink collision. Solitons 1, 3, 5 and 7 are kinks, while 2, 4 and 6 are antikinks. Initial conditions are  $x_1=-x_7=-39.541867403$, $V_1=-V_7=0.1$, $x_2=-x_6=-24.29923$, $V_2=-V_6=0.05$, $x_3=-x_5=-12$, $V_3=-V_5=0.025$, and $x_4=0$, $V_4=0$.}
\label{fig7}
\end{figure*}

Referring to Fig.~\ref{fig7}(a), we note that solitons with odd (even) number are kinks (antikinks). The initial conditions to ensure that all solitons collide at one point are  $x_1=-x_7=-39.541867403$, $V_1=-V_7=0.1$, $x_2=-x_6=-24.29923$, $V_2=-V_6=0.05$, $x_3=-x_5=-12$, $V_3=-V_5=0.025$, and $x_4=0$, $V_4=0$.  Maximal over spatial coordinate energy densities and elastic strains are shown as the functions of time in Fig.~\ref{fig7} (b-f). The extreme parameters of the collision are: 
\begin{equation}\label{umax7}
u^{(7)}_{\max}\approx 100,
\end{equation}
\begin{equation}\label{kmax7}
k^{(7)}_{\max}\approx 33,
\end{equation}
\begin{equation}\label{emax7}
e^{(7)}_{\max}\approx 100,
\end{equation}
\begin{equation}\label{pmax7}
p^{(7)}_{\max}\approx 2,
\end{equation}
\begin{equation}\label{epsmax7}
\varepsilon^{(7)}_{\max}\approx 14, \quad \varepsilon^{(7)}_{\min}\approx -2.7.
\end{equation}

All the results of this Section are shown in Table \ref{cnls_numer_summary}.
 
\begin{table}[pht] \center \begin{small}
\caption{Summary on maximal energy densities and tensile and compressive elastic strains in collision of $N$ solitons.} \centering
\begin{tabular}{ | c | c | c |c | c | c | c | c |}
\hline
$N$                  & $1$   & $2$   & $3$   & $4$   & $5$   & $6$   & $7$    \\
$u_{\max}$           & $4$   & $8$   & $20$   & $32$   & $52$ & $72$   & $100$  \\
$k_{\max}$           & $0$   & $8$   & $5$ & $32$   & $17$ & $72$   & $33$  \\
$e_{\max}$           & $2$ & $2$ & $18.5$ & $10$ & $52$ & $25$ & $100$  \\
$p_{\max}$           & $2$ & $2$& $2$ & $2$ & $2$ & $2$ & $2$ \\
$\varepsilon_{\max}$ & $2$   & $2$   & $6$ & $4.5$ & $10$ & $7$ & $14$  \\
$\varepsilon_{\min}$ & $-2$  & $-2$  & $-2$  & $-4.5$& $-2$  & $-7$& $-2.7$ \\
\hline
\end{tabular}
\label{cnls_numer_summary}
\end{small}
\end{table}

\section {Concluding remarks} \label{Sec:V}

The extreme values of elastic strain and energy in the collision of $N$ slow kinks/antikinks (with $N \leq 7$) in the integrable sG model were calculated numerically. 

The values of elastic strain and energy are calculated when all $N$ kinks/antikinks collide at one point. This happens when the kinks and antikinks approach the collision point alternatively (i.e., no two adjacent solitons before the collision are of the same type). When all $N$ solitons collide at one point, they produce a high energy density spot.

We have separated the total energy density, $u$, into three components, the kinetic energy density, $k$, the elastic strain energy density, $e$, and the potential energy density due to the on-site potential, $p$. Their maximal values observed in the collisions of $N$ kinks/antikinks are also given in Table~\ref{cnls_numer_summary}. We note that $k^{(N)}_{\max}$ increases rapidly with $N$ for even $N$, while for odd $N$ a rapid growth with $N$ is observed for $e^{(N)}_{\max}$. These two energy densities have a dominant contribution to the maximal total energy density.

For many applications, e.g., in the solid state physics, it is important to know the maximal values and the sign of the maximal elastic strain observed in $N$-soliton collisions \cite{Ferro1,Ferro3,Ferro2,Buijnsters,Wrinklon,Yamal,FK}. Large tensile or compressive stress can result in phase transformations \cite{Megumi,Zhilyaev}, or defect nucleation and fracture can happen under tensile stress above the strength limit of the considered medium \cite{Ch2017}. The last two lines of Table~\ref{cnls_numer_summary} contain the information about extreme values of strain in multi-kink collisions. For $N=5$ the maximal tensile strain of $10$ is registered, which is 5 times higher than in the core of a single kink. For $N=7$, maximal tensile strain of $14$ is observed, which is 7 times higher than in the core of a kink. 

For the future works, it is important to calculate the maximal energy density that can be achieved in multi-soliton collisions in other integrable and non-integrable systems of different dimensionality. For example, one can examine similar issues and design such collisions in other Klein-Gordon field theoretical models (e.g. in the $\phi^8$, or $\phi^{12}$ models~\cite{NewJHEP,phi6a,phi6b,147,Khare}), as well as in the nonlinear Schr{\"o}dinger equation \cite{Sh1,Sh2} and in the discrete models free of the Peierls-Nabarro potential \cite{PNPfree}. It would be particularly interesting to explore if the relevant phenomenology persists therein. Next, it would be extremely interesting to search for the physical phenomena that can be related to the high energy density spots or/and highly strained regions generated by multi-soliton collisions. Some of these ideas are under consideration and will be reported in the future works.

\section*{Acknowledgements}
A.M.M thanks the Quchan branch Islamic Azad University for their financial support under the grant. S.V.D. was supported by the grant of the Russian
Science Foundation (No. 16-12-10175). Stay of D.S. at IMSP RAS was partly supported by the Russian Science Foundation, grant No. 14-13-00982.
%

\end{document}